\documentstyle[12pt]{article}
\input tcilatex

\begin{document}

\author{Sergiu Ostaf  \\ 
The Tiraspol State University, Department of Physics,\\ 5 Iablocikin str.,
Chisinau, MD2069, MOLDOVA, chdom@moldnet.md}
\title{{\sc Two-loop Locally Anisotropic Corrections for nonlinear }$\sigma ${\sc %
-model.}}
\maketitle

\section{Abstract}

The article gives the explicit interpretation of the nonlinear locally
anisotropic $\sigma -$model of Boze string\footnote{%
The author is indebted to Dr. Sergiu Vacaru for some conceptual discussions.}%
. The equations of motions and tensor of energy-momentum, the background
field method applied to get order extensions of the effective action
consequently accommodated to the geometry of locally anisotropic spaces,
additional contributions-corrections resulted from the richer geometric
character of the model are calculated and interpreted from the possible
physical point of view. It is shown that the strings theory considered on
spaces with local anisotropy and nonlinear connections are renormalizable
taken the similar nature of the explicitly calculated anisotropic
corrections with the Riemannian one.

Author explains shortly necessary geometrical background and conventions of
locally anisotropic geometry as well as the effects of this geometry on
strings. Mention is given particularly to the interpretation of the results
and ''new'' aspects that features the locally anisotropic strings form their
Riemannian counterparts.

\section{Introduction}

The concept of locally anisotropic strings and of the nonlinear $\sigma -$%
model had already been advanced in some works \cite{VacaruAP, VacaruNP,
ostafconstanta98}. There has been given the substantial body of conceptual
motivation and necessary geometrical language adopted to develop the locally
anisotropic (super)strings \cite{Lovelace1984, Fradkin1985, Callan1985}, and
their extensions to superstrings \cite{Hull1987, Ketov1992, Schwarz1985}.

The diversity of results formulated and obtained in locally anisotropic
physics \cite{vacarumonografie, vsp96, vst96, vcl96, vg} provoked and
justified a special interest for modelling of nonlinear $\sigma -$model on
spaces with local anisotropy.

As the nonlinear geometry considered sometimes inadequate or rather
inappropriate for the easily accepted standard mathematical tools to explain
physical phenomena, physicists always remain ''unhappy'' that the diverse
nature of the phenomenon considered is somewhat straiten out by isotropic
and/or linear approximations. A physicist always wants that linear isotropic
mathematical tool applied to describe a phenomena keep the model flexible
enough to consider possible intrusions deviations, interferences,
interactions and corrections. It has been a predominant practice that
intrinsically desired flexibility is introduced or accounted for with the
help of some external to the initial mathematical tool of model assumptions
or approximations. It does though make happy the physicist, however
demonstrating after all, how innovative the human intelligence is, yet
leaves one with the sense of self-inconsistency of the theory itself.

The proposed example demonstrates how the nonlinear geometry-spaces with
nonlinear connection or spaces with local anisotropy, spaces that evidently
will make happy any physicist to modulate a phenomena, once he/she has an
appropriate mathematical tool to work with, allows the generalization of the
non-linear $\sigma -$model. The explicitly constructed model on locally
anisotropic spaces gives not just necessarily locally anisotropic
corrections, but is manifestly renormalizable and consistent with the
standard previously obtained results of nonlinear $\sigma -$model and what
is important essentially widens the horizon of modelling string theories.

The rich nature of the fiber-base structures allows, apart from construction
of strict physically friendly results, hypothetical interpretation of the
fiber-base objects, as for instance the interaction gauge like fields. This
objects serve also simultaneously as some additional terms-calculated
explicitly in this work-for the standard nonlinear $\sigma -$model. The
fiber-base objects, reasonably judged, may lead to some further developments
of string theory as for, instance the consistent explication of the string
interactions. The above suggested ideas may justifiably present certain
interest for physicists.

The article is structured sa follows: first goes a brief introduction into
locally anisotropic geometry and objects; than the equations of motion and
tensor of energy-momentum of locally anisotropic strings is considered;
follows the accommodation of the standard background field method; later the
effective action order extension and locally anisotropic counterterms
discussed widely; the article finalizes conclusions and discussions of the
obtained results.

\section{An Outline of Locally Anisotropic Geometry.}

We consider $\epsilon =(E,\pi ,F,G,M)$ to be a locally trivial vector
bundle, $v$- bundle, where $F$ is a vector space with dim$F$=$m$, $G$ is a
group of automorphism of $F$, $\pi =E\rightarrow M$ is a surjective map and
a differentiable manifold $F,\dim E$=$m+n$ is called as the total space of $%
v $-bundle $\epsilon .$ We can locally parameterize $\epsilon $ by
coordinates $u^\alpha =(x^i,y^a),$ where $i,j,k,l,m,n,u,...$ take value $%
0,1,...n-1$ and Greek indices $a,b,c,d,e,f,...$ take value $0,1,...m-1.$

Coordinate transforms $(x^i,y^a)\rightarrow (x^{i^{\prime }},y^{a^{^{\prime
}}})$ on differentiable manifold $\epsilon $ are given by formulas $%
x^{i^{^{\prime }}}=x^{i^{^{\prime }}}(x^i),rank(\partial x^{i^{^{\prime
}}}/\partial x^i)=n,$and $y^{a^i}=M_a^{a^{^{\prime
}}}(x)y^a,M_a^{a^{^{\prime }}}(x)\in G$

We provide $\epsilon $ with the structure of nonlinear connection that
splits $v$-bundle into horizontal, $HE$ and vertical $VE$ subbundles of the
tangent bundle $TE$

\begin{equation}
\label{s1}TE=HE\oplus VE 
\end{equation}
For a $N$-connection on $\epsilon $ one can associate the covariant
derivation operator

\begin{equation}
\label{s2}\nabla _XA=y^i\left\{ \frac{\partial A^a}{\partial x^i}%
+N_i^a(x,A)\right\} s_a 
\end{equation}
where $s_a$ are local linearly independent sections of $\epsilon ,A=A^as_a$
and $Y=Y^is_i$ is a vector field decomposition on local bases $s_i$ on $M.$ $%
N_i^a(x,y)$ are called as coefficients of $N$- connection.

The transformation law for $N$-connection under coordinate transforms

\begin{equation}
\label{s3}N_{i^{^{\prime }}}^{a^{^{\prime }}}\frac{\partial x^{i^{^{\prime
}}}}{\partial x^i}=M_a^{a^{^{\prime }}}(x)N_i^a-\frac{\partial
M_a^{a^{^{\prime }}}}{\partial x^i}(x)y^a 
\end{equation}

$N$-connection $N_i^a(x,y)$ is characterized by its curvature

\begin{equation}
\label{s4}\Omega =\frac 12\Omega _{ij}^adx^i\wedge dx^j\otimes \frac
\partial {\partial y^a} 
\end{equation}

with coefficients

\begin{equation}
\label{s5}\Omega _{ij}^a=\frac{\partial N_j^a}{\partial x^i}-\frac{\partial
N_j^a}{\partial x^j}+N_j^b\frac{\partial N_i^a}{\partial y^b}-N_i^b\frac{%
\partial N_j^a}{\partial y^b} 
\end{equation}

For further needs we define a locally adopted (to $N$- connection) reaper
basis as

\begin{equation}
\label{s6}u_\alpha =\frac \delta {\delta u_\alpha }=(x_i=\frac \delta
{\delta x_i}=\frac \partial {\partial x_i}-N_i^a(x,y)\frac \partial
{\partial y^a};y_a=\frac \delta {\delta y^a}=\frac \partial {\partial y^a}) 
\end{equation}

The dual basis to $x_{\alpha \text{ }}$is as

\begin{equation}
\label{s7}u^\alpha =\delta u^\alpha =\left( x^i=dx^i;x^a=\delta
y^a=dy^a+N_i^a(x,y)dx^i\right) 
\end{equation}

The algebra of tensors distinguished fields on $\epsilon $ can be introduced
by using bases

$$
t=t_{j_{1,\ldots }j_qb_{1,\ldots }b_s}^{i_{1,\ldots }i_pa_{1,\ldots
}a_r}(x,y)\frac \delta {\delta x^{i_1}}\otimes \cdots \otimes \frac \delta
{\delta x^{i_r}}\otimes dx^{j_1}\otimes \cdots \otimes dx^{i_p}\otimes
\ldots 
$$

\begin{equation}
\label{s8}\frac \partial {\partial y^{a_1}}\otimes \cdots \otimes \frac
\partial {\partial y^{a_r}}\otimes \cdots \otimes \delta y^{b_1}\otimes
\cdots \otimes \delta y^{b_s} 
\end{equation}

Along with nonlinear $N$-connection one can define a distinguished linear
connection ($d$- connection) $\Gamma _{\beta \gamma }^\alpha $ associated to
a fixed $N$- connection structure on $\epsilon $

\begin{equation}
\label{s9}D_\gamma \frac \delta {\delta u_\beta }=D_{\frac \delta {\delta
u_\gamma }}\frac \delta {\delta u_\beta }=\Gamma _{\beta \gamma }^\alpha
\frac \delta {\delta u_\alpha } 
\end{equation}

Torsion $T_{\beta \gamma \text{ }}^\alpha $and curvature $R_{\beta \text{ }%
}{}^\alpha {}_{\delta \gamma }$ of $d$- connection $\Gamma ^\alpha {}_{\beta
\gamma }$ are defined respectively as

\begin{equation}
\label{s10}T\left( \frac \delta {\delta u^\gamma },\frac \delta {\delta
u^\beta }\right) =T^\alpha {}_{\beta \gamma }\frac \delta {\delta u^\alpha } 
\end{equation}
where $T^\alpha {}_{\beta \gamma }=\Gamma ^\alpha {}_{\beta \gamma }-\Gamma
^\alpha {}_{\gamma \beta }+\varpi ^\alpha {}_{\beta \gamma }$ and

\begin{equation}
\label{s11}R\left( \frac \delta {\delta u^\delta },\frac \delta {\delta
u^\gamma },\frac \delta {\delta u^\beta }\right) =R_\beta {}^\alpha
{}_{\gamma \delta }\frac \delta {\delta u^\gamma } 
\end{equation}
where $R_\beta {}^\alpha {}_{\gamma \delta }=\frac \delta {\delta u^\delta
}\Gamma ^\alpha {}_{\beta \gamma }-\frac \delta {\delta u^\gamma }\Gamma
^\alpha {}_{\beta \delta }+\Gamma ^\alpha {}_{\beta \gamma }\Gamma ^\alpha
{}_{\varphi \delta }-\Gamma ^\varphi {}_{\beta \delta }\Gamma ^\alpha
{}_{\varphi \gamma }+\Gamma ^\alpha {}_{\beta \gamma }\varpi ^\alpha
{}_{\gamma \delta }.$ Throughout the formulas used below $\varpi ^\alpha
{}_{\beta \gamma }$ are nonholonomic coefficients of locally adopted reapers 
$$
\left[ \frac \delta {\delta u^\alpha },\frac \delta {\delta u^\beta }\right]
=\frac \delta {\delta u^\alpha }\frac \delta {\delta u^\beta }-\frac \delta
{\delta u^\beta }\frac \delta {\delta u^\alpha }=\varpi ^\gamma {}_{\alpha
\beta }\frac \delta {\delta u^\gamma }. 
$$

Global decomposition of bundle $\epsilon $ into horizontal and vertical
parts by nonlinear connection structure splits components of $d$-connection
and $d$- tensor fields into horizontal and vertical ones. Locally they
appear for horizontal components: 
$$
D_{\frac \delta {\delta x^k}}^h\left( \frac \delta {\delta x^i}\right)
=L^i{}_{jk}(x,y)\frac \delta {\delta x^i};\text{ }D_{\frac \delta {\delta
x^k}}^h\left( \frac \delta {\delta y^b}\right) =L^a{}_{bk}(x,y)\frac \delta
{\delta y^a} 
$$

and $D_{\frac \delta {\delta x^k}}^hf=\frac{\delta f}{\delta x^k}=\frac{%
\partial f}{\partial x^k}-N^a{}_k(x,y)\frac{\partial f}{\partial y^a}$ where 
$f(x,y)$ is a scalar function on $\epsilon $ and for vertical components 
$$
D_{\frac \delta {\delta y^c}}^v\left( \frac \delta {\delta x^j}\right)
=C^i{}_{jc}(x,y)\frac \delta {\delta x^i};D_{\frac \delta {\delta
y^c}}^v\left( \frac \delta {\delta y^b}\right) =C^a{}_{bc}(x,y)\frac
\partial {\partial y^a},\text{ }D_{\frac \delta {\delta y^c}}^vf=\frac{%
\partial f}{\partial x^k}. 
$$

For components of torsion an explicit calculation gives 
\begin{eqnarray*}
hT\left( \frac \delta {\delta x^k},\frac \delta {\delta x^j}\right)
&=&T^i{}_{jk}\frac \delta {\delta x^i},vT\left( \frac \delta {\delta
x^k},\frac \delta {\delta x^j}\right) =T^a{}_{jk}\frac \partial {\partial
y^a}, \\
hT\left( \frac \partial {\partial y^b},\frac \partial {\partial x^i}\right)
&=&P^i{}_{jb}\frac \delta {\delta x^i},vT\left( \frac \partial {\partial
y^b},\frac \delta {\delta x^j}\right) =P^a{}_{jb}\frac \partial {\partial
y^a}, \\
vT\left( \frac \partial {\partial y^c},\frac \partial {\partial y^b}\right)
&=&S^a{}_{bc}\frac \partial {\partial y^a},
\end{eqnarray*}

the corresponding components of torsion are

\begin{equation}
\label{s12}
\begin{array}{c}
T^i{}_{jk}=L^i{}_{jk}-L^i{}_{kj},T^a{}_{jk}=R^a{}_{jk}=
\frac{\delta N^a{}_k}{\delta x_j},P^i{}_{jb}=C^i{}_{jb}, \\ P^a{}_{jb}=\frac{%
\partial N^a{}_j}{\partial y^b}-L^a{}_{bj},S^a{}_{bc}=C^a{}_{bc}-C^a{}_{cb} 
\end{array}
\end{equation}
For the components of curvature an explicit calculation gives 
\begin{equation}
\label{s13}
\begin{array}{c}
R\left( \frac \delta {\delta x^k},\frac \delta {\delta x^j}\right) \frac
\delta {\delta x^l}=R_l{}^i{}_{jk}\frac \delta {\delta x^i},R\left( \frac
\delta {\delta x^k},\frac \delta {\delta x^j}\right) \frac \partial
{\partial y^b}=R_b{}^a{}_{jk}\frac \partial {\partial y^a}, \\ 
R\left( \frac \partial {\partial y^c},\frac \partial {\partial x^k}\right)
\frac \delta {\delta x^j}=P_j{}^i{}_{kc}\frac \delta {\delta x^i},R\left(
\frac \partial {\partial y^c},\frac \delta {\delta x^k}\right) \frac
\partial {\partial y^b}=P_b{}^a{}_{kc}\frac \partial {\partial y^a}, \\ 
R\left( \frac \partial {\partial y^c},\frac \partial {\partial y^b}\right)
\frac \delta {\delta x^j}=S_j{}^i{}_{bc}\frac \delta {\delta x^i},R\left(
\frac \partial {\partial y^i},\frac \partial {\partial y^c}\right) \frac
\partial {\partial y^b}=S_b{}^a{}_{cd}\frac \partial {\partial y^a} 
\end{array}
\end{equation}

the corresponding components of curvature are

\begin{equation}
\label{s14}
\begin{array}{c}
R_j{}^i{}_{kl}=
\frac{\delta L^i{}_{jk}}{\delta x^l}-\frac{\delta L^i{}_{jl}}{\delta x^k}%
+L^h{}_{jk}L^i{}_{hl}-L^h{}_{jl}L^i{}_{hk}+C^i{}_{ja}R^a{}_{kl}, \\ 
R_b{}^a{}_{kl}=
\frac{\delta L^a{}_{bk}}{\delta x^l}-\frac{\delta L^a{}_{bl}}{\delta x^k}%
+L^c{}_{bk}L^a{}_{cl}-L^c{}_{bl}L^a{}_{ck}+C^a{}_{bc}R^c{}_{kl}, \\ 
P_j{}^i{}_{kl}=
\frac{\partial L^i{}_{jk}}{\partial y^l}-C^i{}_{jc\Vert
k}+C^i{}_{jb}P^b{}_{kl}, \\ P_j{}^i{}_{kc}=
\frac{\partial L^a{}_{bk}}{\partial y^c}-C^a{}_{bc\Vert
k}+C^a{}_{bd}P^d{}_{kc}, \\ S_j{}^i{}_{bc}=
\frac{\partial C^i{}_{jb}}{\partial y^c}-\frac{\partial C^i{}_{jc}}{\partial
y^b}+C^h{}_{jb}C^i{}_{hc}-C^h{}_{jc}C^c{}_{hb}, \\ S_b{}^a{}_{cd}=\frac{%
\partial C^b{}_{da}}{\partial y^c}-\frac{\partial C^a{}_{bd}}{\partial y^c}%
+C^f{}_{bc}C^a{}_{fd}-C^f{}_{bd}C^a{}_{fc} 
\end{array}
\end{equation}

{}In addition to $d$- connection structure, metric structure on $v$- bundle $%
\epsilon $ being associated to a map $G(u):T_u\epsilon \otimes T_u\epsilon
\rightarrow R.$ Choosing a concordance between $N$- connection and metric $G$
on $\epsilon $ when condition $G\left( \frac \delta {\delta x^i},\frac
\partial {\partial y^a}\right) =0$ or equivalently $%
N^a{}_i(x,y)=G_{ib}(x,y)G^{ba}(x,y)$ are held. A metric $G$ on $\epsilon $
is defined by two independent $d$-tensors $g_{ij}(x,y)$ of type $\left( 
\begin{array}{cc}
2 & 0 \\ 
0 & 0 
\end{array}
\right) $ and $h_{ab}(x,y)$ of type$\left( 
\begin{array}{cc}
0 & 0 \\ 
0 & 2 
\end{array}
\right) $ and with respect to the local adapted bases can be written as $%
G=g_{ij}(x,y)dx^i\otimes dx^j+h_{ab}(x,y)\delta y^a\delta y^b.$ $d$-
connection $\Gamma _{\beta \gamma \text{ }}^\alpha $ is compatible with
metric structure $G:$

\begin{equation}
\label{s15}D_\alpha G_{\beta \gamma }=\frac \delta {\delta u^\alpha
}G_{\beta \gamma }-G_{\varphi \gamma }\Gamma ^\varphi {}_{\beta \alpha
}-G_{\beta \varphi }\Gamma ^\varphi {}_{\gamma \alpha }=0 
\end{equation}

In $v$- bundle $\epsilon $ we can consider the canonical (metric) $d$-
connection $\Gamma (N)$ with components $\stackrel{s}{\Gamma }^\alpha
{}_{\beta \gamma }=(L^i{}_{jk},L^a{}_{bi},C^i{}_{jc},C^a{}_{bc})$ determined
by metric $G.$ Below we write out the components of $d$- connections:

\begin{equation}
\label{s16}
\begin{array}{c}
L^i{}_{jk}=\frac 12g^{ip}\left( 
\frac{\delta g_{pj}}{\delta x^k}+\frac{\delta g_{pk}}{\delta x^j}-\frac{%
\delta g_{jk}}{\delta x^p}\right) , \\ L^a{}_{bi}=
\frac{\partial N^a{}_i}{\partial y^b}+\frac 12h^{ac}\left( \frac{\delta
h_{bc}}{\delta x^i}-\frac{\partial N^a{}_i}{\partial y^b}h_{ac}-\frac{%
\partial N^a{}_i}{\partial y^c}h_{ab}\right) , \\ C^i{}_{jb}=\frac 12g^{ik}
\frac{\partial g_{jk}}{\partial y^b}, \\ C^a{}_{bc}=\frac 12h^{ad}\left( 
\frac{\partial h_{bd}}{\partial y^c}+\frac{\partial h_{dc}}{\partial y^b}-%
\frac{\partial h_{bc}}{\partial y^d}\right) 
\end{array}
\end{equation}
The Ricci tensor $R_{\beta \gamma }=R_{\beta \text{ }}{}^\alpha {}_{\gamma
\alpha }$ defines $cd$- connection and has the following components with
respect to adapted reaper basis

\begin{equation}
\label{s16a}
\begin{array}{c}
R_{ij}=R_i{}^k{}_{jk},R_{ia}=P_i{}^k{}_{ak}=-P_{ia}, \\ 
R_{ai}=P_a{}^b{}_{ib}=P_{ai},R_{ab}=S_a{}^c{}_{bc}=S_{ab} 
\end{array}
\end{equation}
In general tensor Ricci is not symmetric. The scalar curvature of $cd$-
connection $R=G^{\alpha \beta }R_{\alpha \beta \text{ }}$is given by $R=R+C,$
where $R=g^{ij}R_{ij}$ and $S=h^{ab}S_{ab.}$

\section{Equation of Motion and Energy-Momentum Tensor of Locally
Anisotropic Strings.}

We are intending to get strings equation of motion considering the movement
of the two-dimensional surface imbed in $(n+m)$ dimensional locally
anisotropic space-time geometry and expansion of the tensor energy-momentum
to Locally Anisotropic Geometry (LAG). We discuss some particular
consequences of the Locally Anisotropic Strings (LAS).

The $\sigma $- model which corresponds to the Bose string action with
Witten-Zumino term in locally anisotropic case is

\begin{equation}
\label{2.1}I_0=\frac 1{4\pi \alpha ^{^{\prime }}}\int dz^\alpha \left(
\delta ^Au^\alpha \delta _Au^\beta G_{\alpha \beta }(u)+i\epsilon
^{AB}\delta _Au^\alpha \delta _Bu^\beta H_{\alpha \beta }(u)\right) 
\end{equation}
Varying with respect to $\frac \delta {\delta u^{\alpha \text{ }}}$ in order
to get string's equation of motion we note that due to the locally
antisymmetric nature of the second term in (\ref{2.1}) it does not
contribute to string's equation of motion and taking into account the
compatibility condition 
\begin{equation}
\label{2.2}D_\gamma G_{\alpha \beta }(u)=\frac \delta {\delta u^\gamma
}G_{\alpha \beta }-\Gamma ^\gamma {}_{\alpha \sigma }G_{\gamma \beta
}-\Gamma ^\gamma {}_{\beta \sigma }G_{\alpha \gamma }=0 
\end{equation}
Lagrangian is given by 
\begin{equation}
\label{2.3}{\cal L}(u,G)=\delta ^Au^\alpha \delta _Au^\beta G_{\alpha \beta
}(u) 
\end{equation}
we get 
$$
\begin{array}{c}
\frac{\delta {\cal L}}{\delta u^\gamma }=\frac \delta {\delta u^\gamma
}(\delta ^Au^\alpha \delta _Au^\beta G_{\alpha \beta }(u))=\frac \delta
{\delta u^\gamma }\left( \frac \partial {\partial z^A}u^\alpha \frac
\partial {\partial z_A}u^\beta \right) G_{\alpha \beta }(u)+\delta
^Au^\alpha \delta _Au^\beta \frac \delta {\delta u^\gamma }G_{\alpha \beta
}(u)= \\ \delta ^Au^\alpha \delta _Au^\beta \frac \delta {\delta u^\gamma
}G_{\alpha \beta }(u)=\frac \partial {\partial z^A}\delta ^\alpha {}_\gamma
\frac \partial {\partial z_A}u^\beta +\frac \partial {\partial z^A}u^\alpha
\frac \partial {\partial z_A}\delta ^\beta {}_\gamma + \\ 
\left( 
\begin{array}{c}
\partial _k-N^a{}_k\partial _a \\ 
\partial _c 
\end{array}
\right) \left( 
\begin{array}{cc}
g_{ij}(u) & 0 \\ 
0 & h_{ab}(u) 
\end{array}
\right) =\delta ^\alpha {}_\gamma \delta ^A\delta _Au^\beta +\left( 
\begin{array}{c}
\delta _kg_{ij} \\ 
\partial _ch_{ab} 
\end{array}
\right) =0 
\end{array}
$$
multiplying on $G^{\alpha \beta \text{ }}$and bearing in mind that $%
G^{\alpha \beta }G_{\alpha \beta }=(m+n-1)$ we have the string equation of
motion of the locally anisotropic string 
\begin{equation}
\label{2.4}(m+n-1)\partial _A\partial _Bu^\beta +\Gamma ^\alpha {}_{\beta
\gamma }(u)\partial _Au^\alpha \partial _Bu^\beta =0 
\end{equation}
where $\Gamma ^\alpha {}_{\beta \gamma }$ as was expected is a symmetric
part of the distinguished linear connection in the adapted bases and the
last becomes possible owing to (\ref{2.2}) .

One can split $\Gamma ^\alpha {}_{\beta \gamma }(u)$ into symmetric part $%
s\Gamma ^\alpha {}_{\beta \gamma }(u)$ and nonsymmetric $n\Gamma ^\alpha
{}_{\beta \gamma }$ parts on the different bases.

Let us consider one of them minding one more that will be taken as essential
in next section.

Symmetric part of $\Gamma ^\alpha {}_{\beta \gamma }(u)$ can be obtained by
simple substraction of nonsymmetric part $n\Gamma ^\alpha {}_{\beta \gamma
}(u)$ of $\Gamma ^\alpha {}_{\beta \gamma }(u)$ : 
\begin{equation}
\label{2.5}s\Gamma ^\alpha {}_{\beta \gamma }(u)=\Gamma ^\alpha {}_{\beta
\gamma }(u)-\left( 
\begin{array}{cc}
0 & L^i{}_{jc} \\ 
C^a{}_{bi} & 0 
\end{array}
\right) =\left( 
\begin{array}{cc}
L^i{}_{jk} & 0 \\ 
0 & C^a{}_{bc} 
\end{array}
\right) 
\end{equation}
$s\Gamma ^\alpha {}_{\beta \gamma }(u)$ Levi-Civita connection or any other.
Than equation (\ref{2.4}) take the form of a system of two equations 
\begin{equation}
\label{2.6}\left\{ 
\begin{array}{c}
(m-1)\partial _A\partial _Bx^j+L^i{}_{jk}(x,y)\partial _Ax^i\partial _Bx^j=0
\\ 
(n-1)\partial _A\partial _By^b+C^a{}_{bc}(x,y)\partial _Ay^a\partial _By^b=0 
\end{array}
\right\} 
\end{equation}
the equation (\ref{2.6}) gives rise to the idea that in locally anisotropic
case we have two way directions to move each being in corresponding
subspaces (fiber and base). However, they are not independent due to
independence of $L^i{}_{jk}(x,y)$ and $C^a{}_{bc}(x,y)$ on $u(x,y)$ so that
motion in fiber bears an impact from the point of view of physical
applications consequences. We intend to discuss it in the next paper.

Our next step is to have energy-momentum tensor extended to the case of the
locally anisotropic geometry. We proceed by varying with respect to $%
G_{\alpha \beta }(u):$

$$
\frac{\delta I_0}{\delta G_{\alpha \beta }(u)}=\frac 12\int d^2z\sqrt{G}%
T_{\alpha \beta }(u)=0 
$$
where 
\begin{equation}
\label{2.7}T_{\alpha \beta }=\delta _Au_\alpha \delta ^Au_\beta -\frac
12G_{\alpha \beta }G^{\gamma \delta }\delta _Au_\beta \delta ^Au_\delta =0. 
\end{equation}
The later result pefectly fits with the standard one in Riemannian case\cite
{poliakov}, as is to expect.

\section{Locally Anisotropic Geometry Background Field Method.}

This section deals with formulating of the Riemannian coordinates to the
case of the locally anisotropic geometry. We consider also the possibility
to obtain symmetric part of $\Gamma ^\alpha {}_{\beta \gamma }(u)$ on the
basis of having certain restrictions fixed on linear connection in the
adapted basis.

One embraces here a different approach adapted in \cite{VacaruNP, VacaruAP}
rather developing the standard normal (Riemannian) coordinates \cite
{rashevsky}, generalizing them to the case of locally anisotropic geometry 
\cite{ostafconstanta98}. Author follows the general logic of the order
extension contained in \cite{Fridling1986, Mukhi1986}. As far as $u^\alpha
=(x^i,y^a),\xi ^\alpha =(\theta ^i,\lambda ^a)$ and making shift $u^\alpha
\rightarrow u^\alpha +\xi ^\alpha $ and expanding it as a power series in $%
\xi ^\alpha $ we get 
\begin{equation}
\label{3.1}u^\alpha =u_0^\alpha +\xi ^\alpha s-\frac 12\left( \stackrel{s}{%
\Gamma }_{\beta \gamma }^\alpha \right) _0\xi ^\beta \xi ^\gamma -\frac
1{3!}\left( \Gamma ^\alpha {}_{\beta \gamma \delta }\right) _0\xi ^\beta \xi
^\gamma \xi ^\delta -\cdots 
\end{equation}
and accommodating the well-known relations to the case 
$$
\begin{array}{c}
\Gamma ^\delta {}_{\left( \alpha _1,\alpha _2\ldots \alpha _n\right)
}(u)=\frac 1N(
\frac{\delta \Gamma ^\delta {}_{\alpha _1,\alpha _2,\ldots \alpha _{n-1}}}{%
\delta u^{\alpha _n}}-\Gamma ^\delta {}_{\gamma ,\alpha _2,\ldots \alpha
_{n-1}}\stackrel{s}{\Gamma }^\delta {}_{\alpha _1\alpha _2}- \\ \Gamma
^\delta {}_{\alpha _1\gamma \alpha _3,\ldots \alpha _{n-1}}\stackrel{s}{%
\Gamma }^\delta {}_{\alpha _2\alpha _n}-\Gamma ^\delta {}_{\alpha _1,\alpha
_2,\ldots \gamma }\stackrel{s}{\Gamma }^\gamma {}_{\alpha _{n-1},\alpha _n)} 
\end{array}
$$
and

\begin{equation}
\label{3.2}
\begin{array}{c}
\Gamma ^\delta {}_{\left( \alpha _1,\alpha _2\ldots \alpha _n\right)
}(u)=\frac 1N( 
\frac{\delta \Gamma ^\delta {}_{\alpha _1,\alpha _2,\ldots \alpha _{n-1}}}{%
\delta u^{\alpha _n}}- \\ (N-1)\Gamma ^\delta {}_{\gamma ,\alpha _2,\ldots
\alpha _{n-1}}\stackrel{s}{\Gamma }^\delta {}_{\alpha _1\alpha _n}) 
\end{array}
\end{equation}
where 
\begin{equation}
\label{3.3}\frac{\delta ^2u^\alpha }{\delta s^2}+\stackrel{s}{\Gamma }%
^\alpha {}_{\beta \gamma }\frac{\delta u^\beta }{\delta s}\frac{\delta
u^\gamma }{\delta s}=0,v^\alpha =\xi ^\alpha s 
\end{equation}
we finally obtain decomposition

\begin{equation}
\label{3.4}u^\alpha =u_0^\alpha +v^\alpha -\frac 12\left( \stackrel{s}{%
\Gamma }_{\beta \gamma }^\alpha \right) _0v^\beta v^\gamma -\frac
1{3!}\left( \Gamma ^\alpha {}_{\beta \gamma \delta }\right) _0v^\beta
v^\gamma v^\delta -\cdots 
\end{equation}
we referred to $\stackrel{s}{\Gamma }^\alpha {}_{\beta \gamma }$ as a
symmetric part of the distinguished linear connection in adapted bases and
bearing in mind (\ref{2.2}) and (\ref{2.4}) takes form 
\begin{equation}
\label{3.5}
\begin{array}{c}
u^\alpha =u_0^\alpha +\xi ^\alpha s-\frac 12\left( 
\stackrel{s}{\Gamma }_{\beta \gamma }^\alpha \right) _0\xi ^\beta \xi
^\gamma s^2- \\ \frac 1{3!}\left( \Gamma ^\alpha {}_{\beta \gamma \delta
}\right) _0\xi ^\beta \xi ^\gamma \xi ^\delta s^3-\cdots 
\end{array}
\end{equation}
and as \cite{rashevsky} 
\begin{eqnarray}
\left\{ \partial _{(\beta }\Gamma _{\alpha )\mu }^\nu \right\} _0 &=&1/3%
\stackrel{0}{R}_{(\alpha \beta )\mu }^\nu ,  \label{s13} \\
\left\{ \partial _{(\gamma \beta }\Gamma _{\alpha )\mu }^\nu \right\} _0
&=&-1/2\stackrel{}{\stackrel{0}{R}_{\mu (\gamma \beta ,\alpha )}^\nu }, 
\nonumber \\
\left\{ \partial _{(\delta \gamma \beta }\Gamma _{\alpha )\mu }^\nu \right\}
_0 &=&-3/5(2/9\stackrel{}{\stackrel{0}{R}_{(\alpha .\beta }^{.\rho .\nu }\stackrel{0}{R}_{\delta .\gamma )\mu }^\omega }\stackrel{0}{G}_{\rho \omega
}+R_{\mu (\delta .\gamma ,\alpha \beta )}^{..\nu };  \nonumber \\
&&\ \ ...  \nonumber
\end{eqnarray}
a locally anisotropic tensor field $W_{\alpha _1...\alpha _p}(u),$ can be
order expanded 
\begin{equation}
\label{s14}W_{\alpha _1...\alpha _p}=\stackrel{0}{W}_{\alpha _1...\alpha
_p}+\left( \frac{\partial W_{\alpha _1...\alpha _p}}{\partial \xi ^\nu }%
\right) _0\xi ^\nu +\frac 1{2!}\left( \frac{\partial ^2W_{\alpha _1...\alpha
_p}}{\partial \xi ^\mu \partial \xi ^\nu }\right) _0\xi ^\mu \xi ^\nu +... 
\end{equation}
one can express $W_{\alpha _1...\alpha _p}$ by means of the (\ref{s13}) and (%
\ref{s14}) 
\begin{eqnarray*}
W_{\alpha _1...\alpha _p} &=&\stackrel{0}{W}_{\alpha _1...\alpha _p}+%
\stackrel{0}{W}_{\alpha _1...\alpha _p,\mu }\xi ^\mu +1/2!\{\stackrel{0}{W}%
_{\alpha _1...\alpha _p,\mu \omega } \\
&&\ \ -1/3\stackrel{p}{\stackunder{k=1}{\sum }}\stackrel{0}{R}_{\mu \alpha
_k\omega }^\nu W_{\alpha _1...\alpha _{k-1}\nu \alpha _{k+1...\alpha
_p}}\}\xi ^\mu \xi ^\omega +1/3!\{\stackrel{0}{W}_{\alpha _1...\alpha _p,\mu
\omega \sigma } \\
&&\ \ -\stackrel{p}{\stackunder{k=1}{\sum }}\stackrel{0}{R}_{\mu \alpha
_k\omega }^\nu W_{\alpha _1...\alpha _{k-1}\nu \alpha _{k+1...\alpha
_p},\sigma } \\
&&\ \ -1/2\stackrel{p}{\stackunder{k=1}{\sum }}\stackrel{0}{R}_{\mu \alpha
_k\omega ,\sigma }^\nu W_{\alpha _1...\alpha _{k-1}\nu \alpha _{k+1...\alpha
_p}}\}\xi ^\mu \xi ^\omega \xi ^\sigma +..
\end{eqnarray*}
The general form is the same as in Riemannian space-time but the essence
consists in the fact that $\xi $ splits in fiber $\lambda ^a$ and $\theta ^i$
components and as we see later contractions are possible only for $\widehat{%
\left( \lambda ^a\lambda ^b\right) }$or $\widehat{\left( \theta ^i\theta
^j\right) }$ as far as dim$M\neq $dim$F$ (dimension of the fiber is not
equal to the dimension of the base in general case. Yet, $\widehat{\left(
\lambda \theta \right) }$ contractions present special interest in vector
bundle space).

Symmetric part of $\Gamma ^\alpha {}_{\beta \gamma }$ can be also achieved
by reducing local anisotropic space-time i.e. restricting the mixed
components of the $d$- connection. Proceeding this way we get that torsion
vanishes and we can define locally and along with a curve $\stackrel{s}{%
\Gamma }^\alpha {}_{\beta \gamma }$ and summing up over the fiber-fiber and
base-base indices in (\ref{s16}) we get that \cite{landau}, 
\begin{equation}
\label{3.6}
\begin{array}{c}
L^a{}_{ai}(u)=\frac \delta {\delta x^i}\ln \left( 
\sqrt{-h}\right) \\ C^i{}_{ia}(u)=\frac \partial {\partial y^a}\ln \left( 
\sqrt{-g}\right) 
\end{array}
\end{equation}
The geometrical sense of the considered restrictions is evident from the
definition of connection components 
\begin{equation}
\label{3.7}
\begin{array}{c}
D_{\frac \delta {\delta x^k}}^h(\frac \partial {\partial
x^a})=L^a{}_{ak}(x,y)\frac \partial {\partial y^a} \\ 
D_{\frac \partial {\partial y^c}}^v(\frac \delta {\delta
x^i})=C^i{}_{ic}(x,y)\frac \delta {\delta x^i} 
\end{array}
\end{equation}
bases vectors $\frac \partial {\partial y^a},\frac \delta {\delta x^i}$ lost
their flexibility in $\frac \delta {\delta x^k},\frac \partial {\partial
y^c} $directions respectively but did not in $\frac \partial {\partial
y^c},\frac \partial {\partial x^k}$ directions. Our assumption makes locally
anisotropic geometry to be rather awkward, but even it allows us to test the
simplest extension of Riemannian bosonic strings to the locally anisotropic
bosonic strings.

The components of the curvature take form: 
\begin{equation}
\label{3.8}
\begin{array}{c}
R_j{}^i{}_{kl}=
\frac{\delta L^i{}_{jk}}{\delta x^l}-\frac{\delta L^i{}_{jl}}{\delta x^k}%
+L^h{}_{jk}L^i{}_{hl}-L^h{}_{jl}L^i{}_{hk}, \\ R_b{}^a{}_{kl}=
\frac{\delta L^a{}_{bk}}{\delta x^l}-\frac{\delta L^a{}_{bl}}{\delta x^k}%
+L^c{}_{bk}L^a{}_{cl}-L^c{}_{bl}L^a{}_{ck}, \\ P_j{}^i{}_{kl}=
\frac{\partial L^i{}_{jk}}{\partial y^c}-C^i{}_{jc\Vert k}, \\ 
P_j{}^i{}_{kc}=
\frac{\partial L^a{}_{bk}}{\partial y^c}-C^a{}_{bc\Vert k}, \\ 
S_j{}^i{}_{bc}=
\frac{\partial C^i{}_{jb}}{\partial y^c}-\frac{\partial C^i{}_{jc}}{\partial
y^b}+C^h{}_{jb}C^i{}_{hc}-C^h{}_{jc}C^c{}_{hb}, \\ S_b{}^a{}_{cd}=\frac{%
\partial C^b{}_{da}}{\partial y^c}-\frac{\partial C^a{}_{bd}}{\partial y^c}%
+C^f{}_{bc}C^a{}_{fd}-C^f{}_{bd}C^a{}_{fc} 
\end{array}
\end{equation}
The number of curvature components are just the same as in the local
anisotropic case that permits us to make a conclusion that reducement will
give a quantitative appropriate result to the original locally anisotropic
case. The reduced $d$- connection in components is 
\begin{equation}
\label{3.9}\stackrel{s}{\Gamma }^\alpha {}_{\beta \gamma
}=(L^i{}_{jk},L^a{}_{bi},C^i{}_{jc},C^a{}_{bc}). 
\end{equation}
The reduced components of Ricci tensor in adapted bases take form 
$$
R_{ij}=R_i{}^k{}_{jk},R_{ia}=-P_i{}^k{}_{ka}=-\stackrel{2}{P}%
_{ia},R_{ai}=P_a{}^b{}_{ib}=\stackrel{1}{P}%
_{ai},R_{ab}=S_a{}^c{}_{bc}=S_{ab}. 
$$

\section{Effective Action Order Extension and Locally Anisotropic
Counterterms}

The effective action of a locally anisotropic $\sigma -$model of Boze string
is given by

\begin{eqnarray}
I &=&\frac 1{4\pi \alpha ^{^{\prime }}}\int d^2z(\sqrt{\eta }\eta
^{AB}\delta ^Au^\alpha \delta _Au^\beta G_{\alpha \beta }(u)+i\epsilon
^{AB}\delta _Au^\alpha \delta _Bu^\beta H_{\alpha \beta }\,(u)+  \nonumber
\label{s4.1} \\
&&\alpha ^{^{\prime }}\sqrt{\eta }_{<i>}R^{(2)}\phi (u))  \label{s19}
\end{eqnarray}
where the second term is a nonsymmetric one $H_{\alpha \beta }\,=-H_{\beta
\alpha }\,$, It's been used the following notations herein :

\begin{equation}
\label{s20}
\begin{array}[t]{c}
D_A\xi ^\alpha =\delta _A\xi ^\alpha +
\stackrel{s}{_{<i>}\Gamma }^\alpha {}_{\beta \gamma }\delta _Au^\alpha \xi
^\beta , \\ \stackrel{s}{_{<i>}\Gamma }^\alpha {}_{\beta \gamma }=\frac
12G^{\alpha \beta }\left( \delta _\gamma G_{\epsilon \beta }+\delta _\beta
G_{\epsilon \gamma }-\delta _\epsilon G_{\beta \gamma }\right) , \\ 
T_{\alpha \beta \gamma }=\frac 12\left( D_\gamma H_{\alpha \beta }+D_\beta
H_{\gamma \alpha }+D_\alpha H_{\beta \gamma }\right) , \\ 
=\frac 12\left( \delta _\gamma H_{\alpha \beta }+\delta _\beta H_{\gamma
\alpha }+\delta _\alpha H_{\beta \gamma }\right) , \\ 
D_\gamma H_{\alpha \beta }=\delta _\gamma H_{\alpha \beta }-H_{\alpha
\epsilon <i>}\Gamma ^\alpha {}_{\beta \gamma }-H_{\epsilon \beta <i>}\Gamma
^\epsilon {}_{\alpha \gamma }, \\ 
\text{ }\delta _\tau =\frac \delta {\delta u^\tau }. 
\end{array}
\end{equation}

A standard order extension with respect to $\xi $ and considering also the
peculiarities of the locally anisotropic space gives \cite{Mukhi1986}. 
\begin{eqnarray}
I[u] &=&I[\overline{u}]+\frac 1{4\pi \alpha ^{^{\prime }}}\{[\frac
12G_{\alpha \beta }(u)\delta _Au^\alpha D^Au^\beta +G_{\alpha \beta
}(u)D_A\xi ^\alpha D^A\xi ^\beta  \nonumber \\
&&\ \ +_{<i>}R_{\alpha \beta \gamma \delta }\delta _Au^\alpha \delta
^Au^\delta \xi ^\beta \xi ^\gamma +\frac 13D_{\alpha <i>}R_{\beta \gamma
\delta \epsilon }\delta _Au^\beta \delta _Bu^\epsilon \xi ^\alpha \xi
^\gamma \xi ^\delta  \nonumber \\
&&\ \ +\frac 43_{<i>}R_{\alpha \beta \gamma \delta }\delta _Au^\delta D_B\xi
^\alpha \xi ^\beta \xi ^\gamma  \nonumber \\
&&\ \ +\frac 1{12}(D_\alpha D_{\beta <i>}R_{\beta \gamma \delta \epsilon
}+4_{<i>}R_{\delta \eta \alpha \text{ }<i>}^\beta R_{\beta \gamma \delta
\epsilon })\delta _Au^\eta \delta _Bu^\epsilon \xi ^\alpha \xi ^\epsilon \xi
^\gamma \xi ^\delta  \nonumber \\
&&\ \ +\frac 12D_{\beta \text{ }<i>}R_{\beta \gamma \delta \epsilon }\delta
_Bu^\epsilon D_A\xi ^\alpha \xi ^\gamma \xi ^\delta +\frac 13_{<i>}R_{\alpha
\gamma \delta \epsilon }D_A\xi ^\alpha D^A\xi ^\epsilon \xi ^\gamma \xi
^\delta ]+  \nonumber \\
&&\ \ +i\varepsilon ^{AB}[\frac 12D_\beta T_{\alpha \delta \gamma }\delta
_Au^\alpha \delta ^Au^\delta \xi ^\gamma \xi ^\beta +T_{\alpha \beta \gamma
}\delta _Au^\alpha D^A\xi ^\beta \xi ^\gamma  \nonumber  \label{s4.5} \\
&&\ \ +\frac 16(D_AD^AT_{\alpha \gamma \delta }+2T_{\alpha \theta \text{ }<i>}^\beta R_{\beta \gamma \delta \epsilon })\delta _Au^\alpha \delta
^Au^\delta \xi ^\alpha \xi ^\gamma \xi ^\delta  \nonumber \\
&&\ \ +\frac 23D_AT_{\alpha \beta \gamma }\delta _Au^\alpha D_B\xi ^\beta
\xi ^\gamma \xi ^\delta +\frac 13T_{\alpha \beta \gamma }D_A\xi ^\alpha
D^A\xi ^\beta \xi ^\gamma  \nonumber \\
&&\ \ +\frac 1{24}(D_\vartheta D_\delta D_\gamma T_{\beta \epsilon \alpha
}+6D_\vartheta T_{\alpha \theta \text{ }<i>}^\beta R_{\beta \gamma \delta
\epsilon }+2T_{\theta \alpha \text{ }}^\beta D_{\vartheta \text{ }<i>}R_{\beta \gamma \delta \epsilon }])  \nonumber \\
&&\ \ \times \delta _Au^\beta \delta _Bu^\epsilon \xi ^\alpha \xi ^\gamma
\xi ^\delta \xi ^\vartheta +\frac 14(D_\gamma D_\delta T_{\theta \gamma
\alpha }+T_{\gamma \alpha <i>}^\beta R_{\beta \gamma \delta \epsilon } 
\nonumber \\
&&\ \ -\frac 13T_{\theta \alpha \text{ }<i>}^\beta R_{\beta \gamma \delta
\epsilon })\times \delta _Au^\theta D_B\xi ^\alpha \xi ^\gamma \xi ^\delta
\xi ^\epsilon .  \label{s4.5} \\
&&\ \ +\frac 14D_\alpha T_{\beta \gamma \delta }D_A\xi ^\alpha D^A\xi ^\beta
\xi ^\gamma \xi ^\delta ]+O(\xi ^5)\}.  \nonumber
\end{eqnarray}

In what follows one explicitly calculates exclusively the contributions
given by the richer aspects of the locally anisotropic apace time only.
Also, we omit the contributions of the contributions of WZW term solely
given the intention to avoid overwhelming of the calculations. The obtained
results will be added to the standard one cited here from \cite{Metsaev}.

So, the additional action $I_{la}[u]$ to be added to the standard action $%
I[u]$ comes explicitly from the properties of the locally anisotropic
strings. The general structure of the locally anisotropic effective action
of non-linear $\sigma -$model is: 
\begin{equation}
\label{s25}I_{la}^G[u]=I_{la}[\overline{u}%
]+I_{la}^g(x)+I_{la}^h(y)+I_{la}^{g-h}(x,y); 
\end{equation}
where $I_{la}^g(x)$ string effective action on base coinciding to standard
Riemannian one. The term $I_{la}^{g-h}(x,y)$ has influence of both fiber and
base. The contributions of $I_{la}^{g-h}(x,y)$ projected on base, so that to
the standard model, add nonlinear anisotropic corrections we pursue to
calculate. The contribution of $I_{la}^{g-h}(x,y)$ can be virtually
interpreted\footnote{%
The nature of the interaction, to be interpreted from the structure of the
additional terms, will be considered as we go along with presentation.} as
the effective action of interaction between effective action of, say base
residing string $I_{la}^g(x)$ and similarly fiber string $I_{la}^h(y).$ 
\begin{eqnarray*}
I_{la}^{g-h}(x,y) &=&(4\pi \alpha ^{^{\prime }})^{-1}\{D_A\theta
^iD^A\lambda ^a+R_{abkl}\delta _Ay^a\delta ^Ax^l\lambda ^b\theta ^k \\
&&\ \ +P_{jikc}\delta _Ax^j\delta ^Ay^c\theta ^i\theta ^k+P_{abkc}\delta
_Ay^a\delta ^Ay^c\lambda ^b\theta ^k+S_{jkbc}\delta _Ax^j\delta ^Ay^c\theta
^k\lambda ^b \\
&&\ \ +\frac 13D_iR_{abkl}\delta _Ay^a\delta _Bx^l\theta ^i\lambda ^b\theta
^k+\frac 13D_dR_{abkl}\delta _Ay^a\delta _Bx^l\lambda ^d\lambda ^b\theta ^k
\end{eqnarray*}

\begin{eqnarray*}
&&+\frac 13D_fP_{jikl}\delta _Ax^j\delta _Bx^l\theta ^i\theta ^k\theta
^f+\frac 13D_dP_{jikl}\delta _Ax^j\delta _Bx^l\theta ^i\theta ^k\lambda ^d \\
&&+\frac 13D_lP_{bakc}\delta _Ay^b\delta _By^c\lambda ^a\theta ^l\theta
^k+\frac 13D_dP_{bakc}\delta _Ay^b\delta _By^c\lambda ^d\lambda ^a\theta ^k
\\
&&+\frac 13D_dS_{jibc}\delta _Ax^j\delta _By^c\lambda ^d\theta ^i\lambda
^b+\frac 13D_lS_{jibc}\delta _Ax^j\delta _By^c\theta ^l\theta ^i\lambda ^b \\
&&+\frac 43R_{bakl}\delta _Ay^bD_B\theta ^l\theta ^k\lambda ^a+\frac
43P_{jikc}\delta _Ax^jD_B\lambda ^c\theta ^k\theta ^i \\
&&+\frac 43P_{bakd}\delta _Ay^bD_B\lambda ^d\theta ^k\lambda ^a+\frac
43S_{jibc}\delta _Ax^jD_B\lambda ^c\lambda ^b\theta ^i
\end{eqnarray*}

\begin{eqnarray*}
&&\ +\frac 1{12}(D_iD_jR_{bakl}+4R_{akl}^cR_{cjib})\delta _Ay^b\delta
_Bx^l\lambda ^a\theta ^k\theta ^i\theta ^j \\
&&\ +\frac 1{12}(D_cD_dR_{bakl}+4R_{akl}^cS_{cdcb})\delta _Ay^b\delta
_Bx^l\lambda ^a\theta ^k\lambda ^c\lambda ^d \\
&&\ +\frac 1{12}(D_mD_nP_{likc}+4P_{ikc}^fR_{fnml})\delta _Ay^c\delta
_Bx^l\theta ^i\theta ^k\theta ^m\theta ^n \\
&&\ +\frac 1{12}(D_cD_dP_{likc}+4P_{ikl}^fR_{fdcb})\delta _Ay^b\delta
_Bx^l\theta ^i\theta ^k\lambda ^c\lambda ^d \\
&&\ +\frac 1{12}(D_iD_jP_{abkc}+4P_{bka\text{ }}^dR_{djic})\delta
_Ay^a\delta _By^c\lambda ^b\theta ^k\theta ^i\theta ^j \\
&&\ +\frac 1{12}(D_dD_eP_{abkc}+4P_{bka\text{ }}^dR_{djic})\delta
_Ay^a\delta _By^c\lambda ^b\theta ^k\lambda ^e\lambda ^d
\end{eqnarray*}

\begin{eqnarray*}
&&\ \ \ \ \ +\frac 1{12}(D_kD_lS_{jibc}+4S_{ibj\text{ }}^dR_{dlkc})\delta
_Ax^j\delta _By^c\theta ^i\lambda ^b\theta ^k\theta ^l \\
&&\ \ \ \ \ +\frac 1{12}(D_aD_eS_{jibc}+4S_{ibj\text{ }}^dR_{dlkc})\delta
_Ax^j\delta _By^c\theta ^i\lambda ^b\lambda ^a\lambda ^e
\end{eqnarray*}
\begin{eqnarray*}
&&\ \ \ \ \ +\frac 12D_jR_{bakl}\delta _By^bD_A\theta ^l\lambda ^a\theta
^j\theta ^k+\frac 12D_dR_{bakl}\delta _By^bD_A\theta ^l\lambda ^a\lambda
^d\theta ^k \\
&&\ \ \ \ \ +\frac 12D_lP_{jikd}\delta _Bx^jD_A\theta ^i\theta ^l\theta
^k\lambda ^d+\frac 12D_aP_{jikd}\delta _By^aD_A\theta ^i\theta ^l\theta
^k\lambda ^d \\
&&\ \ \ \ \ +\frac 12D_lP_{bakd}\delta _By^bD_A\theta ^l\theta ^k\lambda
^d+\frac 12D_cP_{bakd}\delta _By^bD_A\lambda ^c\theta ^k\lambda ^d \\
&&\ \ \ \ \ +\frac 12D_lS_{jibc}\delta _Bx^jD_A\theta ^l\theta ^i\lambda
^b\lambda ^c+\frac 12D_aS_{jibc}\delta _Bx^jD_A\lambda ^a\theta ^i\lambda
^b\lambda ^c
\end{eqnarray*}
\begin{eqnarray}
&&\ \ \ \ +\frac 13R_{bakl}D_A\lambda ^bD^A\theta ^l\lambda ^a\theta
^k+\frac 13P_{jikd}D_A\theta ^jD^A\lambda ^d\theta ^k\lambda ^d  \nonumber \\
&&\ \ \ \ +\frac 13P_{bakc}D_A\lambda ^bD^A\lambda ^c\lambda ^a\theta
^k+\frac 13S_{jibc}D_A\theta ^jD^A\lambda ^c\theta ^i\lambda ^b\};
\label{s26}
\end{eqnarray}
where $I_{la}^h(y)$ the fiber effective action gets quite already familiar
expression 
\begin{eqnarray}
I_{la}^h(y) &=&\frac 1{4\pi \alpha ^{^{\prime }}}\{D_A\lambda ^aD^A\lambda
^b+h_{ab}D_A\lambda ^aD^A\lambda ^b+S_{abcd}\delta _Ay^a\delta ^Ay^d\lambda
^c\lambda ^b  \nonumber \\
&&\ \ \ \ +\frac 13D_eS_{bacd}\delta _Ay^b\delta _By^d\lambda ^e\lambda
^a\lambda ^c+\frac 43S_{bacd}\delta _Ay^bD_B\lambda ^d\lambda ^a\lambda ^c 
\nonumber \\
&&\ \ \ \ +\frac 1{12}(D_eD_uS_{bacd}+4S_{acb\text{ }}^sS_{seud})\delta
_Ay^b\delta _By^d\lambda ^a\lambda ^c\lambda ^e\lambda ^u  \nonumber \\
&&\ \ \ \ +\frac 12D_eS_{bacd}\delta _By^bD_A\lambda ^e\lambda ^d\lambda
^c\lambda ^a+\frac 13S_{bacd}D_A\lambda ^bD^A\lambda ^d\lambda ^a\lambda
^c\};  \label{s27}
\end{eqnarray}
The contribution and effects of $I_{la}^h(y)$ are absolutely symmetric to $%
I_{la}^g(x)$ and to Riemannian string. Therefore, we intentionally omit
consideration in these respect also for the diagrams and counterterms
resulted from $I_{la}^h(y)$ do not interfere with those of base component as
well as, it has been intended to see the effects upon the base evolvements.

Here one can envisage at least two possible ways to count the effects of $%
I_{la}^{g-h}(x,y)$. First is to project the additional terms on base, so
that they lost vehemently the fiber aspect whatsoever. Second, is to
interpret the fiber-base nature of the corrections as they have significance
from the point of view of fiber-base interactions. There are both
conveniently approached, whereas the first one is of primarily interest in
the framework of exclusive calculation of locally anisotropic corrections,
while the second one appealed to justifying the physically charged nature of
the additional terms.

The explicit expression for fiber-base locally anisotropic effective action $%
I_{la}^{g-h}(x,y)$ is given by: 
\begin{eqnarray*}
I_{la}^{g-h}(x,y) &=&[P_{jikc}\delta _Ax^j\delta ^Ay^c+\frac
13D_iR_{abkl}\delta _Ay^a\delta _Bx^l\lambda ^b \\
&&+\frac 13D_dP_{jikl}\delta _Ax^j\delta _Bx^l\lambda ^d+\frac
13D_iP_{bakc}\delta _Ay^b\delta _By^c\lambda ^a \\
&&+\frac 13D_kS_{jibc}\delta _Ax^j\delta _By^c\lambda ^b+\frac
43P_{jikc}\delta _Ax^jD_B\lambda ^c \\
&&+\frac 1{12}(D_cD_dP_{likc}+4P_{ikl}^fR_{fdcb})\delta _Ay^b\delta
_Bx^l\lambda ^c\lambda ^d]\times \theta ^i\theta ^k \\
&&+[\frac 43R_{bakl}\delta _Ay^b\lambda ^a+\frac 12D_dR_{bakl}\delta
_By^b\lambda ^a\lambda ^d+\frac 12D_lP_{bakd}\delta _By^b\lambda ^d \\
&&+\frac 13S_{lkbc}D^A\lambda ^c\lambda ^b+\frac 12D_lS_{jkbc}\delta
_Bx^j\lambda ^b\lambda ^c+\frac 13R_{bakl}D_A\lambda ^b\lambda ^a]\times
D^A\theta ^l\theta ^k \\
&&+\frac 13P_{jikd}\lambda ^d\lambda ^dD_A\theta ^jD^A\theta ^k
\end{eqnarray*}

\begin{eqnarray}
&&\ \ +[\frac 13D_lP_{jikl}\delta _Ax^j\delta _Bx^l+\frac
1{12}(D_iD_lR_{bakl}+4R_{akl}^cR_{clib})\delta _Ay^b\delta _Bx^l\lambda ^a 
\nonumber \\
&&\ \ +\frac 1{12}(D_iD_lP_{abkc}+4P_{bka\text{ }}^dR_{dlic})\delta
_Ay^a\delta _By^c\lambda ^b  \nonumber \\
&&\ \ +\frac 1{12}(D_kD_lS_{jibc}+4S_{ibj\text{ }}^dR_{dlkc})\delta
_Ax^j\delta _By^c\lambda ^b]\times \theta ^i\theta ^k\theta ^l  \nonumber \\
&&\ \ +[\frac 12D_kR_{baji}\delta _By^b\lambda ^a+\frac 12D_lP_{jikd}\delta
_Bx^j\lambda ^d  \nonumber \\
&&\ \ +\frac 12D_aP_{jikd}\delta _By^a\lambda ^d]\times D_A\theta ^i\theta
^l\theta ^k  \nonumber \\
&&\ \ +\frac 1{12}(D_mD_nP_{likc}+4P_{ikc}^fR_{fnml})\delta _Ay^c\delta
_Bx^l\theta ^i\theta ^k\theta ^m\theta ^n  \label{s28}
\end{eqnarray}
the standard procedure for doing calculations of divergencies \cite
{Fridling1986} implies the following notations: 
\begin{eqnarray*}
A_{\varepsilon \gamma \zeta }\times (\xi \xi \xi ) &=&-\frac 16(\eta
^{AB}+\varepsilon ^{AB})(D_\zeta \overline{_{<i>}R}_{\alpha \gamma \beta
\varepsilon }-2\overline{_{<i>}R}_{\alpha \gamma \beta }^\kappa T_{\beta
\varepsilon \kappa })D_Au^\alpha \partial _Bu^\beta \\
&&\times (\xi ^\varepsilon \xi ^\gamma \xi ^\zeta ), \\
B_{\beta \gamma \varepsilon }^B\times (D\xi \xi \xi ) &=&-\frac 23(\eta ^{AB}%
\overline{_{<i>}R}_{\gamma (\alpha \beta )\varepsilon }+\varepsilon ^{AB}%
\overline{_{<i>}R}_{\gamma (\alpha \beta )\varepsilon })\partial _Au^\alpha
\times (D_B\xi ^\beta \xi ^\gamma \xi ^\varepsilon ), \\
C_{\alpha \gamma }^{AB}\times (D\xi D\xi \xi ) &=&-\frac 13\varepsilon
^{AB}T_{\alpha \beta \gamma }\times (\xi ^\alpha D_A\xi ^\alpha D_B\xi
^\gamma ), \\
L_{AB}^{\alpha \beta \gamma \delta }\times (D\xi D\xi \xi \xi ) &=&-\frac
14[\frac 23\eta ^{AB}(\overline{_{<i>}R}_{\alpha \beta }+T_{\alpha \delta
}^\kappa T_{\epsilon \beta \kappa })+\varepsilon ^{AB}\overline{_{<i>}R}%
_{\alpha \delta \epsilon \beta }] \\
&&\times (D_A\xi ^\alpha D_B\xi ^\beta \xi ^\epsilon \xi ^\delta ), \\
H^{\gamma \delta \epsilon \lambda }\times (\xi \xi \xi \xi ) &=&\{-\frac
1{24}(\eta ^{AB}+\varepsilon ^{AB})[D_\epsilon D_\lambda \overline{_{<i>}R}%
_{\alpha \gamma \delta \beta } \\
&&+_{<i>}\overline{R}_{\alpha \gamma \delta }^\kappa \overline{_{<i>}R}%
_{\beta \epsilon \lambda \kappa }+3\overline{_{<i>}R}_{\alpha \gamma \delta
}^\kappa \overline{_{<i>}R}+4D_\lambda \overline{_{<i>}R}_{\alpha \gamma
\delta \kappa }T_{\beta \lambda } \\
&&+4\overline{_{<i>}R}_{\alpha \gamma \delta \kappa }T_{\epsilon \nu
}^\kappa T_{\beta \lambda }^\nu ]\delta _Au^\alpha \delta _Bu^\beta \}\times
(\xi ^\gamma \xi ^\delta \xi ^\epsilon \xi ^\lambda ), \\
X^{\beta \gamma }\times (\xi \xi ) &=&-\frac 12(\eta ^{AB}+\varepsilon ^{AB})%
\overline{_{<i>}R}_{\alpha \beta \gamma \delta }\delta _Au^\alpha \delta
^Au^\delta \times (\xi ^\beta \xi ^\gamma ).
\end{eqnarray*}

The counterterms coming out of the one loop diagrams are proportional to

\begin{equation}
\label{s29}
\begin{array}{r}
X_{\alpha \beta }=
\stackunder{\gamma \delta }{\sum }\frac 12\left( _{<i>}R_{\alpha \delta
\gamma \beta }\delta _Au^\alpha \delta ^Au^\beta \xi ^\gamma \xi ^\delta
-\frac 12i\epsilon ^{AB}D_\delta T_{\alpha \beta \gamma }\delta _Au^\alpha
\delta _Bu^\beta \xi ^\delta \xi ^\gamma \right) \\ +1/3m_{<i>}^2R_{\alpha
\beta }u^\alpha u^\beta \\ 
=\frac 12(_{<i>}R_{\alpha \beta }-1/4T_{\alpha \beta }^2)-\frac 12i\epsilon
^{AB}D_\delta T_{\alpha \beta \delta }+\frac 13m_{<i>}^2R_{\alpha \beta
}u^\alpha u^\beta . 
\end{array}
\end{equation}
where there has been considered the contributions of the relations from base 
\cite{Metsaev} and of corrections fiber-base projected on base (\ref{s28}). 
\begin{equation}
\label{s30}
\begin{array}{c}
I_{count,1}^{(la)}=\Gamma ^{(1)}{}_{\infty ,1}=(4\pi \epsilon )^{-1}\int
d^az(\delta ^Au^\alpha \delta _Au^\beta (_{<i>}R_{\alpha \beta }-\frac
14(T_{\alpha \beta })^2 \\ 
-\epsilon ^{AB}\delta _Au^\alpha \delta _Bu^\beta (\frac 12D_\lambda
T_{\alpha \beta \lambda })+\frac 13m_{<i>}^2R_{\alpha \beta }u^\alpha
u^\beta ) 
\end{array}
\end{equation}
In the explicit form the results projected on base (\ref{s28}) are given: 
\begin{eqnarray}
I_{count,1}^{(la)} &=&(4\pi \epsilon )^{-1}\int d^az(\delta ^Ax^i\delta
_Ax^j(R_{ij}-1/4T_{ij}^4)  \nonumber \\
&&\ \ \ \ +V_{jB}\delta _Ax^j+K_{ij}\delta _Ax^i\delta _Bx^j+W_{iA}\delta
_Bx^i  \nonumber \\
&&\ \ \ \ +\epsilon ^{AB}\delta _Ax^i\delta _Bx^j(-\frac 12D_\lambda
T_{ij\lambda })+\frac 13m^2R_{ij}x^ix^j)  \label{s31}
\end{eqnarray}
Expression (\ref{s31}) gives counterterms with one loop contributions
resulted from (\ref{s25}), where one introduced the notations 
\begin{eqnarray}
F_{lA} &=&1/3D_kR_{kabl}\delta _Ay^a\lambda ^b,\text{ }M_{jB}=1/3D_kS_{kjbc}%
\delta _By^c\lambda ^b,\text{ }K_{jl}=4/3D_dP_{jl}\lambda ^d  \label{s31a} \\
C_{lA} &=&\frac 1{12}(D_cD_dP_{lc}+4P_l^fR_{fdcb})\delta _Ay^b\lambda
^c\lambda ^d,\text{ }O_{jB}=P_{jc}D_B\lambda ^c,U_{jA}=P_{jc}\delta ^Ay^c, 
\nonumber \\
W_{jA} &=&U_{jA}+C_{jA.},V_{jB}=M_{jB}+O_{jB}.  \nonumber
\end{eqnarray}

Expression (\ref{s31}) differs from the standard effective action by the
component 
\begin{equation}
\label{s32}pI_{count,1}^{(g-h)}=(4\pi \epsilon )^{-1}\int d^az(V_{jB}\delta
_Ax^j+K_{ij}\delta _Ax^i\delta _Bx^j+W_{iA}\delta _Bx^i) 
\end{equation}
where fields $V_j^{AB}=\epsilon ^{AB}V_{jB}\delta _Ax^j,$ $%
W_i^{AB}=W_{iA}\delta _Bx^i$ are transformed as gauge like fields, so that
the only possible invariant of which has four dimension and cannot
contribute to UV divergencies of two dimensional space. Therefore, the only
additional UV divergent contribution remains 
\begin{equation}
\label{s33}pI_{count,1}^{(g-h)}=(4\pi \epsilon )^{-1}\int d^az(K_{ij}\delta
_Ax^i\delta _Bx^j) 
\end{equation}
The last produces a set of new diagrams, presented below. The structure of $%
pI_{count,1}^{(g-h)}$ has not been essentially different from the rest of
the standard counterterms that thus allows to presuppose, within the first
mentioned above approach, simple locally anisotropic correction.\footnote{%
See also relevant discussions in S.Vacaru, Locally Anisotropic Interactions:
I. Non-linear Connections in Higher Order Anisotropic Superspaces, E-print:
hep-th/9607194
\par
Locally Anisotropic Interactions: II. Torsions and Curvatures of Higher
Order Anisotropic Superspaces, E-print: hep-th/9607194,
\par
Locally Anisotropic Interactions: III. Higher Order Anisotropic
Supergravity, E-print: hep-th/9607196.}. 
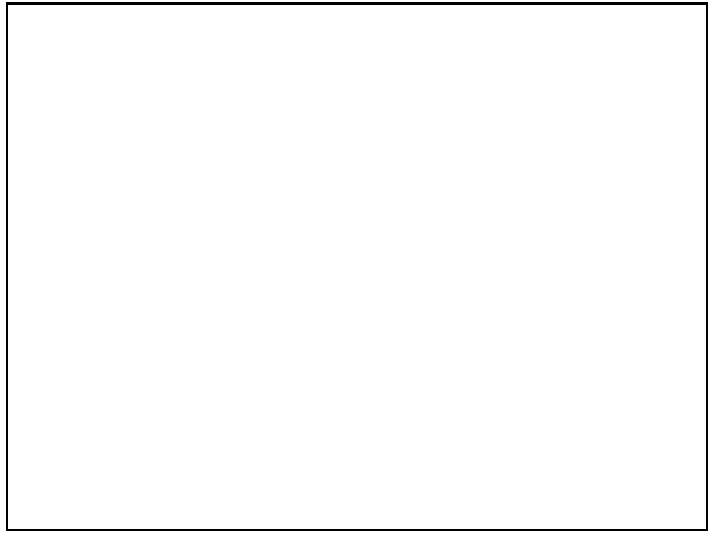
$$
\FRAME{itbpxFX}{3.6608in}{2.7518in}{0pt}{}{}{1loop1-2.bmp}{\special{language
"Scientific Word";type "GRAPHIC";maintain-aspect-ratio TRUE;display
"USEDEF";valid_file "F";width 3.6608in;height 2.7518in;depth 0pt;cropleft
"0";croptop "1";cropright "1";cropbottom "0";filename
'C:/TEZA/NEW/1LOOP1-2.BMP';file-properties "XNPEU";}} 
$$
The contributions of some additional locally anisotropic one loop diagrams
are calculated making use the standard procedure \cite{Alvarez1981}:

\begin{eqnarray}
(1a) &=&-\frac 1{4\varepsilon }K_{ij},  \nonumber \\
(1b) &=&\frac 1{4\varepsilon ^2}(1+\varepsilon )K_{il}D_kH_j^{kl},  \nonumber
\\
(1c) &=&-\frac 1{24\varepsilon }R^{nk}K_{inl}K_{kj}^l,  \nonumber \\
(1d) &=&-\frac 1{4\varepsilon ^2}K_{il}R_j^l,...  \label{s34}
\end{eqnarray}
If not projected on base the diagrams resulted from (\ref{s33}), will bear
tensor sings of fiber. Also the one loop base-fiber diagrams will not be
transformed in two loop diagrams, one of which will be base and another one
fiber loop. This is explicable by the fact that the fiber tensor order is
low enough to prevent order expansion in those one loop counterterms that
contribute to base standard counterterms. The fiber terms allowing fiber
order expansion and forming up fiber one loop do not result in base loops.

At the same time the tensor structure of the terms $V_{jB}\delta _Ax^j,$ $%
W_{iA}\delta _Bx^i$ suggests their gauge like character. Their role is not
interpreted and clear yet.

the action the two loop diagrams to be read from is 
\begin{equation}
\label{s35}%
I_{count,2}=I_{la}^g(x)+pI_{la}^{g-h}(x,y)+pI_{count,2}^{(g-h)}+1/3m^2R_{ij}%
\theta ^i\theta ^j+H-dependentpart 
\end{equation}
where 
\begin{eqnarray}
pI_{count,2}^{(g-h)} &=&+(4\pi \epsilon )^{-1}\int
d^az(R_{ij}-1/4T_{ij}^2+K_{ij})D_A\theta ^iD^A\theta ^j  \nonumber \\
&&\ \ +[1/2D_iD_j(R_{lk}-1/4T_{lk}^2+K_{lk})  \nonumber \\
&&+R_{ijl}^k(R_{lk}-1/4T_{lk}^2+K_{lk})]\times \delta _Ax^j\delta
_Bx^i\theta ^i\theta ^j \\
&&\ \ +2D_f(R_{ij}-1/4T_{ij}^2+K_{ij})\delta _Ax^jD_B\theta ^i\theta ^f 
\nonumber \\
&&\ \ +D_jV_{Bi}D_A\theta ^i\theta ^j+D_jW_{Ai}D_B\theta ^i\theta ^j 
\nonumber \\
&&+[1/3R_{ijk}^l(V_{Bl}\delta _Ax^k+W_{Al}\delta _Bx^k) \\
&&\ \ +1/2D_iD_j(V_{Bl}\delta _Ax^k+W_{Al}\delta _Bx^k) \\
&&-1/3R_{ikj}^l(V_{Bl}\delta _Ax^k+W_{Al}\delta _Bx^k)]\times \theta
^i\theta ^j  \label{s36}
\end{eqnarray}

Below we bring only some of the two loop diagrams resulted from (\ref{s35}),
which are being projected to base again: 
$$
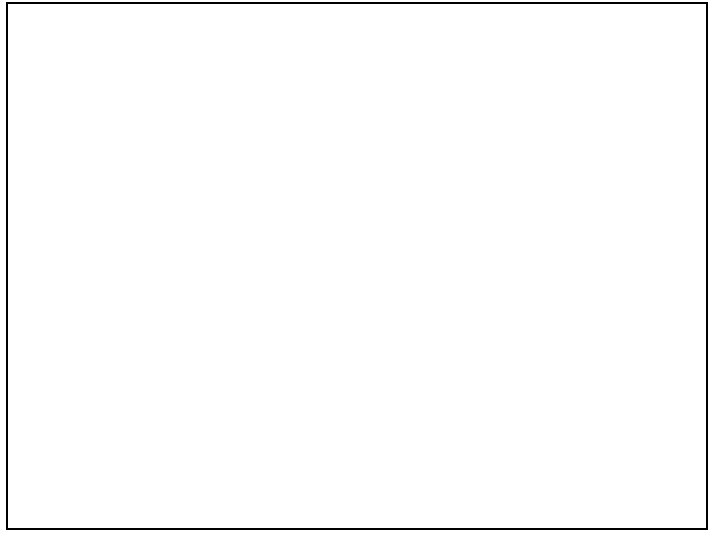
\FRAME{itbpxFX}{3.6608in}{2.7518in}{0pt}{}{}{2loop6a.bmp}{\special{language
"Scientific Word";type "GRAPHIC";maintain-aspect-ratio TRUE;display
"USEDEF";valid_file "F";width 3.6608in;height 2.7518in;depth 0pt;cropleft
"0";croptop "1";cropright "1";cropbottom "0";filename
'C:/TEZA/NEW/2LOOP6A.BMP';file-properties "XNPEU";}} 
$$

The calculations of these two loop locally anisotropic diagrams gives the
result: 
\begin{eqnarray}
(2a) &=&\frac 1{8\pi m^2\varepsilon }K_{klm}^jK_i^{klm},  \nonumber \\
(2b) &=&-\frac 9{16}(\frac 1{4\pi ^2\varepsilon ^2}+\frac 1{2\pi \varepsilon
})K_{iklj}T_{km}T_{lm}  \nonumber \\
(2c) &=&\frac 1{16\pi m^2\varepsilon }K_{iklm}K_j^{klm}  \label{s37}
\end{eqnarray}
Adding up easily the contributions of one and two loop locally anisotropic
counterterms of two loop locally anisotropic $\beta _{(2)ij}^{(g-h)}$
function, one gets the complete two loop locally anisotropic $\beta
_{(2)\alpha \beta }^{(G)la}$ function including all terms of base
fiber-base: 
\begin{equation}
\label{s38}\beta _{(2)\alpha \beta }^{(G)la}=\beta _{(2)ij}^{(g)}+\beta
_{(2)ij}^{(g-h)} 
\end{equation}
where $\beta _{(2)ij}^{(g-h)}$ accounts for the contributions of the
diagrams (\ref{s34}) si (\ref{s37}) and other possible diagrams to be
resulted from locally anisotropic dependence nature of the locally
anisotropic string model. The result (\ref{s38}) taken the already well
known, see for instance \cite{Metsaev}, $\beta _{(2)ij}^{(g)}$ functions
render the explicit contribution projected on base of the effects from
base-fiber, otherwise essentially anisotropic nature of the geometrical
background of the model.

\section{Conclusions and Interpretation of Results}

It is clear that the locally anisotropic spaces, spaces provided with the
structure of nonlinear connection allows conceptual and consistent
generalization of the (super)strings models, as they are modelled on a
explicitly anisotropic geometric background.

The above presented results show that the basic techniques developed in
(super)strings theory after some accommodations to the features of locally
anisotropic spaces are successfully applicable. The locally anisotropic
nonlinear $\sigma -$model is apparently renormalizable, the locally
anisotropic corrections having manifestly similar structure to those of
standard nonlinear $\sigma -$model. The extension of the explicit locally
anisotropic strings to locally anisotropic superstrings is far only a matter
of technical aspects.

An interesting feature of the locally anisotropic strings comes from the
role of the $V_{jB}\delta _Ax^j,$ $W_{iA}\delta _Bx^i$ gauge like fields and
their fiber symmetric counterparts. The explicit form of these fields are
given by (\ref{s31a}). These terms on one hand can contribute to two loop
base diagrams and are gauge like fields in base one loop approximation. At
the same time they are manifestly gauge like fields in fiber space. This
twofold nature of the terms: a contributing term on base and simultaneously
gauge like field in fiber may virtually inspire to physically interesting
interpretations, as for instance can be string interaction \footnote{%
S. Ostaf, Locally Anisotropic Geometric Interpretation of String
Interactions. (in preparation)}.

In the number of above cited works the geometry of locally anisotropic
spaces had been groundly considered physically natural and quite appropriate
bases to modeling and investigating fundamental interactions. The explicitly
presented example is considered by author to be one of the convincing
argument for raising the validity and trustworthy of the locally anisotropic
spaces in physicists' perception.

\end{document}